\documentclass[12pt]{article}
\pdfoutput = 1
\usepackage{amsmath,amssymb,bm,epsfig,afterpage}
\usepackage{mathtools}
\usepackage{cite}
\usepackage{here}
\usepackage{color}
\usepackage{tabularx}
\usepackage{simplewick} 
\usepackage{bbm}

\usepackage{comment}

\renewcommand{\thefootnote}{\fnsymbol{footnote}}
\newcommand{\vev}[1]{{\langle{#1}\rangle}}

\newcommand{\abs}[1]{\left|{#1}\right|}

\newcommand{\eps}{\epsilon}

\newcommand{\order}[1]{\mathcal{O}\left({#1}\right)}
\newcommand{\Ykr}[2]{Y^{({#1})}_{#2}} 

\newcommand{\pr}{\prime}

\numberwithin{equation}{section}

\allowdisplaybreaks[3]
\newcolumntype{Y}{&gt;{\centering\arraybackslash}X} 

\usepackage{hyperref}

\begin{document}

\begin{titlepage}

\begin{flushright}
 {\tt
CTPU-PTC-23-27 \\
EPHOU-23-012  
}
\end{flushright}

\vspace{1.2cm}
\begin{center}
{\Large
{\bf
Fermion Hierarchies in $SU(5)$ Grand Unification 
\\ from $\Gamma_6^\prime$ Modular Flavor Symmetry  
}
}
\vskip 2cm
Yoshihiko Abe$^{\ a}$~\footnote{yabe3@wisc.edu}, 
Tetsutaro Higaki$^{\ b}$~\footnote{thigaki@rk.phys.keio.ac.jp}, 
Junichiro Kawamura$^{\ c}$~\footnote{junkmura13@gmail.com}
and
Tatsuo Kobayashi$^{\ d}$~\footnote{kobayashi@particle.sci.hokudai.ac.jp}

\vskip 0.5cm

{\it $^a$
Department of Physics, University of Wisconsin, Madison, WI 53706, USA
}\\[3pt]

{\it $^b$
Department of Physics, Keio University, Yokohama, 223-8522, Japan
}\\[3pt]

{\it $^c$
Center for Theoretical Physics of the Universe, Institute for Basic Science (IBS),
Daejeon 34051, Korea
}\\[3pt]

{\it $^d$
Department of Physics, Hokkaido University, Sapporo 060-0810, Japan}\\[3pt]

\vskip 1.5cm

\begin{abstract}
We construct a model in which the hierarchies of the quark and lepton masses and mixing 
are explained by the $\Gamma_6^\prime$ modular flavor symmetry.
The hierarchies are realized by the Froggatt-Nielsen-like mechanism 
due to the residual $Z^T_6$ symmetry, approximately unbroken at $\tau \sim i\infty.$
We argue that the $\Gamma_6^{(\prime)}$ symmetry is the minimal possibility 
to realize the up-type quark mass hierarchies, since the Yukawa matrix is symmetric. 
We find a combination of the representations and modular weights
and then show numerical values of $\mathcal{O}(1)$ coefficients 
for the realistic fermion hierarchies.  
\end{abstract}
\end{center}
\end{titlepage}

\clearpage

\renewcommand{\thefootnote}{\arabic{footnote}}
\setcounter{footnote}{0}

\newcommand{\la}{{\lambda}}
\newcommand{\ka}{{\kappa}}
\newcommand{\mQ}{{m^2_{\tilde{Q}}}}
\newcommand{\mU}{{m^2_{\tilde{u}}}}
\newcommand{\mD}{{m^2_{\tilde{d}}}}
\newcommand{\mL}{{m^2_{\tilde{L}}}}
\newcommand{\mE}{{m^2_{\tilde{e}}}}
\newcommand{\mhu}{{m^2_{H_u}}}
\newcommand{\mhd}{{m^2_{H_d}}}
\newcommand{\ms}{{m^2_S}} 
\newcommand{\Ala}{{A_\lambda}}
\newcommand{\Aka}{{A_\kappa}} 

\newcommand{\id}[1]{\mathbf{1}_{#1}}
\newcommand{\ol}[1]{\overline{#1}}
\newcommand{\Lcal}{\mathcal{L}}
\newcommand{\Mcal}{\mathcal{M}}
\newcommand{\Ncal}{\mathcal{N}}
\newcommand{\Ycal}{\mathcal{Y}}
\newcommand{\sg}{\sigma}
\newcommand{\sgb}{\overline{\sigma}}
\newcommand{\del}{\partial}

\newcommand{\htm}{\hat{m}}
\newcommand{\tU}{\widetilde{U}}
\newcommand{\SL}[2]{\mathrm{SL}({#1},{#2})}
\newcommand{\natN}{\mathbb{N}}
\newcommand{\intZ}{\mathbb{Z}}
\newcommand{\oGam}{\overline{\Gamma}} 
\newcommand{\Gam}{{\Gamma}} 
\newcommand{\Ita}{\mathrm{Im}\,\tau}

\newcommand{\CKM}{\mathrm{CKM}}
\newcommand{\SM}{\mathrm{SM}}
\newcommand{\hY}{\hat{Y}}

\newcommand{\nop}[1]{\textcolor{blue}{#1}}

\newcommand{\YA}[1]{{\color{darkviolet} [\textbf{YA:} #1]}}
\definecolor{darkviolet}{rgb}{0.58, 0.0, 0.83}
\newcommand{\violed}[1]{{\color{darkviolet}#1}}
\newcommand{\magenta}[1]{{\color{magenta}#1}}
\def\Re{\mathop{\mathrm{Re}}}
\def\Im{\mathop{\mathrm{Im}}}
\newcommand{\mr}{\mathrm}

\section{Introduction} 

The Grand Unified Theory (GUT) is an attractive framework 
to understand the gauge structure of the Standard Model (SM)~\cite{Georgi:1974sy}.   
In the $SU(5)$ GUT, 
the quarks and leptons in a generation are unified into $\ol{5}$ and $10$ representations. 
As a result, for instance, 
the exactly opposite charges of the electron and the proton 
are manifestly explained.  
The $SU(5)$ GUT predicts the unification of the three gauge coupling constants 
which is consistent with the Minimal Supersymmetric SM (MSSM)~\cite{Dimopoulos:1981yj,Marciano:1981un,Einhorn:1981sx,Ellis:1990zq,Ellis:1990wk,Amaldi:1991cn,Langacker:1991an,Giunti:1991ta}.   
Due to the unification of the quarks and leptons, 
the Yukawa couplings are also constrained to be consistent 
with the $SU(5)$ gauge symmetry, as we shall study in this paper.

The modular flavor symmetry provides an interesting possibility to understand 
the flavor structures of the three generations of quarks and leptons in the SM~\cite{Feruglio:2017spp}. 
Under the modular symmetry, 
Yukawa coupling constants are treated as the so-called modular forms, 
holomorphic functions of modulus $\tau$. 
The finite modular groups $\Gamma_N$, $N\in\natN$, are generalization 
of the non-Abelian discrete flavor symmetries~\cite{deAdelhartToorop:2011re}, 
as well studied in the literature~\cite{Feruglio:2017spp,Kobayashi:2018vbk,Penedo:2018nmg,Novichkov:2018nkm,Ding:2019xna,Liu:2019khw,Novichkov:2020eep,Liu:2020akv,Liu:2020msy,Altarelli:2010gt,Ishimori:2010au,Ishimori:2012zz,Hernandez:2012ra,King:2013eh,King:2014nza,Tanimoto:2015nfa,King:2017guk,Petcov:2017ggy,Feruglio:2019ybq,Kobayashi:2022moq}.     
In fact, we can find isomorphisms  
$\Gamma_2 \simeq S_3$, $\Gamma_3 \simeq A_4$, $\Gamma_4 \simeq S_4$, 
and so on. 
There have been many attempts to understand the flavor structures of the SM 
by the finite modular flavor symmetries~\cite{Criado:2018thu,Kobayashi:2018scp,Ding:2019zxk,Novichkov:2018ovf,Kobayashi:2019mna,Wang:2019ovr,Chen:2020udk,deMedeirosVarzielas:2019cyj,Asaka:2019vev,Asaka:2020tmo,deAnda:2018ecu,Kobayashi:2019rzp,Novichkov:2018yse,Kobayashi:2018wkl,Okada:2018yrn,Okada:2019uoy,Nomura:2019jxj,Okada:2019xqk,Nomura:2019yft,Nomura:2019lnr,Criado:2019tzk,King:2019vhv,Ding:2019gof,deMedeirosVarzielas:2020kji,Zhang:2019ngf,Nomura:2019xsb,Kobayashi:2019gtp,Lu:2019vgm,Wang:2019xbo,King:2020qaj,Abbas:2020qzc,Okada:2020oxh,Okada:2020dmb,Ding:2020yen,Okada:2020rjb,Okada:2020ukr,Nagao:2020azf,Wang:2020lxk,Okada:2020brs,Yao:2020qyy,Kuranaga:2021ujd}.

There are the hierarchies in the masses of the SM fermions 
and in the Cabibbo-Kobayashi-Maskawa (CKM) matrix. 
The residual $\intZ_N$ symmetry of the modular symmetry 
induces hierarchies of Yukawa couplings 
if the modulus $\tau$ has a value near a fixed point. 
For instance, if $\tau \sim i\infty$, 
the $\intZ_N$ symmetry remains in the modular $\Gamma_N$ symmetry, 
and then the Yukawa couplings are suppressed by powers 
of $q_N := e^{ -2\pi \mathrm{Im}\tau/N} \ll 1$ 
depending on the charge under the $\intZ_N$ symmetry. 
This is a realization of the Froggatt-Nielson (FN) mechanism, 
where the powers of the flavon field are replaced by those of $q_N$. 
Recently, this idea is applied for the quark sector based on the modular 
$A_4$~\cite{Petcov:2022fjf,Petcov:2023vws}, $A_4\times A_4 \times A_4$~\cite{Kikuchi:2023jap}, 
$S_4$~\cite{Abe:2023ilq} and $\Gamma_6$~\cite{Kikuchi:2023cap} symmetries,  
and the lepton sector is also studied in Refs.~\cite{Feruglio:2021dte,Novichkov:2021evw}. 
In Ref.~\cite{Abe:2023qmr}, 
the present authors constructed a model for both quark and lepton sectors 
based on $\Gamma_4^\prime \simeq S_4^\prime$ symmetry,  
where $\Gamma_N^\prime$ is the double covering of $\Gamma_N$ symmetry.

In this work, 
we construct a model to explain the hierarchies 
of the quarks and leptons in the $SU(5)$ GUT. 
It will be turned out that 
the residual symmetry should be $\intZ_N$ with $N\ge 5$,  
to realize the hierarchy of the up-type quark masses 
because the up Yukawa matrix is a symmetric matrix. 
Since the $\Gamma_5^{(\prime)}$ symmetry does not have proper representation 
to realize the hierarchy~\cite{Novichkov:2018nkm,Wang:2020lxk} as explained later, 
we shall consider the $\Gamma_6^{\prime}$ symmetry as the minimal possibility. 
We consider $\Gamma_6^\prime$, double covering of $\Gamma_6$, 
to construct a model with non-singlet representations. 
We also study the neutrino sector assuming the type-I seesaw mechanism 
by introducing three $SU(5)$ singlets.

This paper is organized as follows. 
In Sec.~\ref{sec-fhinGUT}, 
we briefly review the flavor structures of the $SU(5)$ GUT and the modular flavor symmetry, 
and then discuss how to explain the hierarchies of the masses and mixing. 
The explicit model is studied and our results of numerical calculations 
for the charged fermions are shown in Sec.~\ref{sec-model}, 
and then the neutrino sector is discussed in Sec.~\ref{sec-neut}. 
Section~\ref{sec-summary} concludes. 
The details of the modular $\Gamma_6^\prime$ symmetry 
are shown in App.~\ref{app-mod6}.

\section{Fermion hierarchies in $SU(5)$ GUT}  
\label{sec-fhinGUT}
\subsection{$SU(5)$ GUT} 

The superpotential at the dimension-4 is given by  
\begin{align}
\label{eq-Wdim4}
 W_4 = \frac{1}{2} y_{ij} \eps_{ABCDE} 10^{AB}_i  10^{CD}_j H^E  
           + h_{ij} 10^{AB}_{i} \ol{5}_{jA} \ol{H}_B,  
\end{align}
where $A,B,\cdots = 1,2,3,4,5$ are the $SU(5)$ indices,  
and $i,j = 1,2,3$ are the flavor indices. 
The MSSM fields are in the $SU(5)$ multiplets as 
\begin{align}
\ol{5}_a = \ol{d}_a, 
\quad 
\ol{5}_{3+\alpha}  =  \eps_{\alpha\beta}\ell^\beta 
\quad 
  10^{a,3+\alpha}       = Q^{a\alpha}, 
\quad 
  10^{ab}               =  \eps^{abc} \ol{u}_c, 
\quad 
  10^{3+\alpha,3+\beta} = \eps^{\alpha\beta} \ol{e}, 
\end{align}
where $a,b,c = 1,2,3$ and $\alpha,\beta = 1,2$ 
are the $SU(3)_C$ and $SU(2)_L$ indices, respectively. 
In these expressions, $\eps$'s with the indices are the complete anti-symmetric tensors. 
The Higgs doublets are included 
in the (anti-)fundamental representations $H$ ($\ol{H}$) as 
\begin{align}
H^{3+\alpha} = \eps^{\alpha\beta} H_{u\beta} , 
\quad 
\ol{H}_{3+\alpha} =  H_{d\alpha}.  
\end{align}
We assume that the triplets are so heavy 
that it will not induce too fast proton decay by a certain mechanism~\cite{Hisano:1992jj,Lucas:1996bc,Goto:1998qg,Murayama:2001ur}~\footnote{
As far as the triplet Higgses are at the GUT scale, 
the proton lifetime would be long enough if soft SUSY breaking parameters are larger than $\order{100}$~TeV~\cite{Hisano:2013exa}.
}.  
Since the Yukawa matrices of the down quarks and charged leptons 
are given by the common Yukawa matrix $h_{ij}$, 
the dimension-4 Yukawa couplings in Eq.~\eqref{eq-Wdim4}  
can not explain the realistic fermion masses.

In this work, we assume that the $SU(5)$ gauge symmetry is broken 
by an adjoint field $\Sigma_A^B$ whose VEV is given by 
\begin{align}
 \vev{\Sigma} = v_\Sigma \times \mathrm{diag}(2,2,2,-3,-3), 
\end{align} 
so that the GUT symmetry is broken down to the SM gauge symmetry. 
We shall consider the dimension-5 interaction involving the adjoint field,    
\begin{align}
 W_5 \ni &\ 
\frac{1}{\Lambda} 
C_{ij} 10^{AB}_i \Sigma_A^C  \ol{5}_{jC} \ol{H}_B,  
\end{align}
which splits the Yukawa couplings of the down quarks and the charged leptons. 
Here $\Lambda$ is a cutoff scale. 
The dimension-5 operators inserting the adjoint field 
to other places will not change the flavor structure,  
and thus we do not write them explicitly.

The Yukawa couplings for the MSSM fields are given by 
\begin{align}
 W = \ol{u} Y_u Q H_u + \ol{d} Y_d H_d + \ol{e} Y_e \ell H_d,  
\end{align}
where the Yukawa matrices are given by 
\begin{align}
\label{eq-YudeSU5}
 \left[Y_u\right]_{ij} = y_{ij},
\quad  
 \left[Y_d\right]_{ij} = h_{ij} + 2 R_\Sigma C_{ij}, 
\quad 
 \left[Y_e\right]_{ij} = h_{ij} - 3 R_\Sigma C_{ij},  
\end{align}
where $R_\Sigma := v_\Sigma/\Lambda$. 
Thus the $\order{R_\Sigma}$ splitting of the down-type Yukawa couplings 
is induced via the VEV of the adjoint scalar field $v_\Sigma$.

\subsection{Modular flavor symmetry}

We consider the so-called principal congruence subgroups $\Gamma(N)$, $N\in\natN$, 
defined as 
\begin{align}
 \Gamma(N):= \left\{ 
\begin{pmatrix}
 a & b \\ c & d 
\end{pmatrix}  
\in SL(2,\intZ), 
\quad 
\begin{pmatrix}
 a & b \\ c & d 
\end{pmatrix}  
\equiv
\begin{pmatrix}
 1 & 0 \\ 0 & 1 
\end{pmatrix} 
\quad 
\mathrm{mod}~N
\right\}, 
\end{align} 
where $\Gamma := SL(2,\intZ) = \Gamma(1)$ 
is the special linear group of $2\times 2$ matrices of integers. 
The modulus $\tau$ is transformed by the group $\Gamma$ as 
\begin{align}
 \tau \to \frac{a\tau+b}{c\tau+d}.  
\end{align}
The finite modular group $\Gamma_N^\prime$ 
is defined as a quotient group $\Gamma_N^\prime := \Gamma/\Gamma(N)$, 
which can be generated by the generators, 
\begin{align}
 S = 
\begin{pmatrix}
 0 & 1 \\ -1 & 0 
\end{pmatrix}, 
\quad 
 T = 
\begin{pmatrix}
 1 & 1 \\ 0 & 1 
\end{pmatrix}, 
\quad 
R = 
\begin{pmatrix}
 -1 & 0 \\ 0 & -1
\end{pmatrix}. 
\end{align} 
One can consider the quotient group $\Gamma_N := \ol{\Gamma}/\Gamma(N)$, 
where  $\ol{\Gamma}:= \Gamma/\intZ^R_2$  
with $\intZ^R_2$ being the $\intZ_2$ symmetry generated by $R$. 
$\Gamma^\prime_N$ is the double covering of the group $\Gamma_N$. 
For $N < 6$, the generators of $\Gamma_N^\prime$ satisfy 
\begin{align}
\label{eq-algSTR}
 S^2 = R, \quad (ST)^3 = R^2 = T^N = 1, 
\quad TR = RT,
\end{align}
and those for $\Gamma_N$ are given by taking $R=1$. 
At $N=6$, the generators should also satisfy 
\begin{align}
 ST^2 ST^3 ST^4 ST^3  = 1,  
\end{align}
in addition to those in Eq.~\eqref{eq-algSTR}. 
Under $\Gamma_N^\prime$, 
a modular form, a holomorphic function of $\tau$, 
$\Ykr{k}{r}$ with representation $r$ and modular weight $k$ 
transforms as 
\begin{align}
 \Ykr{k}{r} \to (c\tau+d)^k \rho(r) \Ykr{k}{r}(\tau),  
\end{align}
where $\rho(r)$ is the representation matrix of $r$. 
We assume that the chiral superfield $\Phi$ with representation $r_\Phi$ 
and weight $-k_\Phi$ transform as 
\begin{align}
  \Phi \to (c\tau+d)^{-k_\Phi} \rho(r_\Phi) \Phi.  
\end{align}
If the modulus is stabilized near a fixed point 
$\tau \sim i$, $w := e^{2\pi i/3}$ or $i\infty$, 
there remains a residual symmetry $\intZ^S_4$, $\intZ^{ST}_3$ or $\intZ^T_N$, respectively.
Here, the superscript represents a generator associated with the residual symmetry.

The modular invariant K\"ahler potential for the kinetic term is given by 
\begin{align}
\label{eq-Kahler}
 K \ni \frac{\Phi^\dag \Phi}{(-i\tau +i\tau^*)^{k_\Phi}},
\end{align}
and hence the canonically normalized chiral superfield is given by 
\begin{align}
\label{eq-cannom}
 \hat{\Phi} = \left(\sqrt{2\mathrm{Im}\tau}\right)^{-k_\Phi} \Phi.  
\end{align}
The normalization factors from $\sqrt{2~\mathrm{Im}\tau}$ 
will be important for $\tau \sim i\infty$, as we shall consider in this work.

\subsection{Assignments of representations} 
 
The hierarchical Yukawa matrices can be realized by the FN mechanism 
with a $\mathbb{Z}_N$ symmetry~\cite{Froggatt:1978nt,Higaki:2019ojq} 
if the modulus field is stabilized near a fixed point.  
We shall consider the FN-like mechanism by $\mathbb{Z}^T_N$ symmetry, and
define $\eps := \sqrt{3} e^{2\pi i \tau/6}$ in Eq.~\eqref{eq-epsdef} throughout this paper. See Appendix~\ref{app-mod6} for more details. 
Since the up-type Yukawa matrix is symmetric in the $SU(5)$ GUT, 
the texture of $Y_u$ is, in general, given by 
\begin{align}
 Y_u \sim 
\begin{pmatrix}
 \eps^{2n} & \eps^{n+m} & \eps^n \\ 
 \eps^{n+m}& \eps^{2m} & \eps^m \\ 
 \eps^{n}  & \eps^{m} & 1 \\ 
\end{pmatrix}, 
\end{align}
where $\eps \ll 1$ and $n, m\in\mathbb{N}$.
The singular values are read as $(\eps^{2n}, \eps^{2m}, 1)$, 
and hence the minimal possibility to obtain the hierarchal up quark masses 
is $(m,n) = (1,2)$, so that the singular values are given by $(\eps^4, \eps^2, 1)$.  
Thus $N>4$ is required to realize the hierarchy by a $\mathbb{Z}_N$ symmetry.

The $\Gamma_6^{(\prime)}$ symmetry is the minimal possibility 
for the up quark mass hierarchy.  
$N>4$ can be realized only from the residual $T$ symmetry 
in the modular $\Gamma_N^{(\prime)}$ symmetry with $N>4$,
where $\mathbb{Z}^T_N$ symmetry remains unbroken at $\tau \sim i\infty$. 
In the case of $N=5$, however, there are no such representations whose $T$-charges  
are $(0,-1,-2) \equiv (0,4,3)$ mod $N=5$~\cite{Novichkov:2018nkm,Wang:2020lxk}. 
Therefore, 
$N=6$ is the minimal possibility to realize the hierarchal up Yukawa couplings.

As shown in detail in Appendix~\ref{app-mod6}, 
the irreducible representations of $\Gamma_6$ are~\cite{Li:2021buv} 
\begin{align}
 1^s_b,\  2_b,\  3^s,\  6,        
\end{align}
where $s=0,1$ and $b=0,1,2$. 
Since $\Gamma_6$ is isomorphic to $\Gamma_2\times \Gamma_3 \simeq S_3\times A_4$, 
upper (lower) indices can be considered as $S_3$ ($A_4$) indices. 
For the double covering $\Gamma_6^\prime$, 
there are additional representations, 
\begin{align}
 2^s_b, \  4_b.   
\end{align}
The charges under the residual $\intZ^T_6$ symmetry 
of the representations less than three dimensions are given by 
\begin{align}
 T(1^s_b) = 3s+2b,
\quad 
 T(2_b)=   
\begin{pmatrix}
 2b \\ 2b+3
\end{pmatrix}, 
\quad 
 T(3^s)=  
\begin{pmatrix}
 3s \\ 3s+2  \\ 3s+4 
\end{pmatrix}, 
\quad 
T(2^s_b) =   
\begin{pmatrix}
 3s + 2b  \\ 3s + 2b +2 
\end{pmatrix}, 
\end{align}
modulo $6$.

At $N=6$, the texture can be realized for,   
\begin{align}
 10_{i=1,2,3} = 1^{0}_1 \oplus 2^1_2 =: 10_1\oplus (10_2, 10_3).
\end{align}
The $\mathbb{Z}^T_6$-charges are $(2, 1, 3)$.  
Since the doublet representation $2^1_2$ exists only for $\Gamma_6^\pr$, 
we should consider the double covering of $\Gamma_6$. 
We can replace $2^1_2$ to $1^1_2\oplus 1^1_0$ which exist in $\Gamma_6$, 
but we do not consider this case because the model is less predictive 
and may need larger modular weights
to have certain modular forms~\cite{Kikuchi:2023cap}.

Since the doublet lepton $\ell_i$ is in $\ol{5}_i$, 
the hierarchal structures of the neutrino masses and the PMNS matrix 
are determined by the representation of $\ol{5}_i$. 
As $\eps\sim (m_u/m_t)^{1/4} \sim \order{0.05}$ 
for the charged fermion hierarchies, 
$\ell_i$ should have a common $\mathbb{Z}^T_6$ charge, 
so that there is no hierarchy in the neutrino sector. 
Thus we assign 
\begin{align}
 \ol{5}_{i=1,2,3} = 1^s_b \oplus  1^s_b \oplus  1^s_b
 =: \ol{5}_1\oplus  \ol{5}_2 \oplus  \ol{5}_3 ,  
\end{align}
with $s=0,1$ and $b=0,1,2$. 
The values of $(s,b)$ will be determined to have the modular forms 
for a given modular weights. 
In this case, the singular values of the down quarks and charged leptons
are given by $\eps^q (\eps^2, \eps, 1)$, 
where $q = 0,1,2,3$ depends on $(s,b)$.  
The textures of the CKM and PMNS matrices are given by 
\begin{align}
 V_{\mathrm{CKM}} \sim 
\begin{pmatrix}
 1 & \eps & \eps^2 \\ 
 \eps & 1 & \eps   \\
\eps^2 & \eps & 1   
\end{pmatrix}, 
\quad
V_{\mathrm{PMNS}} \sim 
\begin{pmatrix}
 1 & 1 & 1 \\ 
  1 & 1 & 1 \\ 
   1 & 1 & 1 \\ 
\end{pmatrix},   
\end{align}
and there is no hierarchy in the neutrino masses originated from $\eps$. 
We discuss the neutrino sector explicitly in Sec.~\ref{sec-neut}. 
We also note that the textures will also be modified 
by powers of $\sqrt{2\mathrm{Im}\tau}$ as in Eq.~\eqref{eq-cannom} 
depending on modular weights, as will be discussed in the next section in which we will take $\sqrt{2\mathrm{Im}\tau}\sim 2.5$.

\section{Model}
\label{sec-model}

\subsection{Yukawa couplings}

We denote the modular weights of $10_{1,2}$ and $\ol{5}_{1,2,3}$ as 
\begin{align}
k_{10} = (k_{10}^1, k_{10}^2), 
\quad 
k_5    = (k_{5}^1,k_{5}^2,k_{5}^3).  
\end{align}
The modular invariant superpotential of the Yukawa couplings are given by 
\begin{align}
 W =&\   \alpha_1 \left( \Ykr{2k_{10}^1}{1_1^0} 10_1 10_1 \right)  H
     +   \alpha_2 \left( \Ykr{k_{10}^1+k_{10}^2}{2^1_2} 10_1 10_2 \right)_1 H
     +   \alpha_3 \left( \Ykr{2k_{10}^2}{3^0} \ol{10}_2 10_2 \right)_1 H
\\ \notag 
   &\ + \sum_{i=1,2,3} \left[
          \left(\beta_{i1}+\frac{\Sigma}{\Lambda} \gamma_{i1}\right)
         \left( \Ykr{k_{10}^1+k_5^i}{1^{s}_{2-b}} \ol{5}_i 10_1 \right) 
        +
          \left(\beta_{i2}+\frac{\Sigma}{\Lambda} \gamma_{i2}\right)
         \left( \Ykr{k_{10}^2+k_5^i}{2^{s+1}_{-b}}  \ol{5}_i 10_2 \right)_1
         \right]  \ol{H},  
\end{align}
where the contractions of the $SU(5)$ indices are implicit. 
It is assumed that $\Sigma,~H$ and $\ol{H}$ are trivial singlets with modular weight $k=0$.
The symbol $(\cdots)_1$ indicates that the trivial singlet combinations 
inside the parenthesis. 
Here $\alpha_i$, $\beta_{1i}$ and $\beta_{2i}$ are $\order{1}$ coefficients. 
The Yukawa matrices for the MSSM quarks and leptons are read as 
\begin{align}
 Y_u =&\ \frac{1}{\sqrt{6}}
\begin{pmatrix}
\sqrt{6}\alpha_1 \Ykr{2k_{10}^1}{1^0_1} & 
  \sqrt{3}\alpha_{2}\left[\Ykr{k_{10}^1+k_{10}^2}{2^1_2}\right]_2 & 
- \sqrt{3}\alpha_{2}\left[\Ykr{k_{10}^1+k_{10}^2}{2^1_2}\right]_1  \\  
 \sqrt{3} \alpha_{2} \left[\Ykr{k^2_{10}+k^1_{10}}{2^1_2}\right]_2 & 
 -\sqrt{2} \alpha_3 \left[Y_{3^0}^{(2k_{10}^2)}\right]_3 & 
   \alpha_3\left[\Ykr{2k^{2}_{10}}{3^0}\right]_2  \\  
-\sqrt{3} \alpha_{2}\left[\Ykr{k^2_{10}+k^1_{10}}{2^1_2}\right]_1  & 
\alpha_3\left[\Ykr{2k^{2}_{10}}{3^0}\right]_2 & 
 \sqrt{2}\alpha_3 \left[\Ykr{2k^2_{10}}{3^0}\right]_1  \\ 
\end{pmatrix}, \\ 
 Y_d  =&\ \frac{1}{\sqrt{2}} 
\begin{pmatrix}
 \sqrt{2} \beta_{11}^d \Ykr{k_5^1+k_{10}^1}{1^{s+1}_{2-b}} & 
 \beta_{12}^d \left[\Ykr{k_5^1+k_{10}^2}{2^{s}_{-b}}\right]_2 & 
-\beta_{12}^d \left[\Ykr{k_5^1+k_{10}^2}{2^{s}_{-b}}\right]_1  
 \\ 
\sqrt{2} \beta_{21}^d \Ykr{k_5^2+k_{10}^1}{1^{s}_{2-b}} & 
 \beta_{22}^d \left[\Ykr{k_5^2+k_{10}^2}{2^{s+1}_{-b}}\right]_2 & 
-\beta_{22}^d \left[\Ykr{k_5^2+k_{10}^2}{2^{s+1}_{-b}}\right]_1  
 \\ 
 \sqrt{2} \beta_{31}^d \Ykr{k_5^3+k_{10}^1}{1^{s}_{2-b}} & 
 \beta_{32}^d \left[\Ykr{k_5^3+k_{10}^2}{2^{s+1}_{-b}}\right]_2 & 
-\beta_{32}^d \left[\Ykr{k_5^3+k_{10}^2}{2^{s+1}_{-b}}\right]_1 & 
 \\ 
\end{pmatrix},  \\
 Y_e  =&\ \frac{1}{\sqrt{2}} 
\begin{pmatrix}
 \sqrt{2} \beta_{11}^e \Ykr{k_5^1+k_{10}^1}{1^{s}_{2-b}} & 
 \sqrt{2} \beta_{21}^e \Ykr{k_5^2+k_{10}^1}{1^{s}_{2-b}} & 
 \sqrt{2} \beta_{31}^e \Ykr{k_5^3+k_{10}^1}{1^{s}_{2-b}} 
 \\ 
 \beta_{12}^e \left[\Ykr{k_5^1+k_{10}^2}{2^{s+1}_{-b}}\right]_2 & 
 \beta_{22}^e \left[\Ykr{k_5^2+k_{10}^2}{2^{s+1}_{-b}}\right]_2 & 
 \beta_{32}^e \left[\Ykr{k_5^3+k_{10}^2}{2^{s+1}_{-b}}\right]_2  
 \\ 
-\beta_{12}^e \left[\Ykr{k_5^1+k_{10}^2}{2^{s+1}_{-b}}\right]_1  &
-\beta_{22}^e \left[\Ykr{k_5^2+k_{10}^2}{2^{s+1}_{-b}}\right]_1  &
-\beta_{32}^e \left[\Ykr{k_5^3+k_{10}^2}{2^{s+1}_{-b}}\right]_1  
 \\ 
\end{pmatrix},  
\end{align}
where 
\begin{align}
 \beta^d_{ij} := \beta_{ij} + 2R_\Sigma \gamma_{ij}, 
\quad 
 \beta^e_{ij} := \beta_{ij} - 3R_\Sigma \gamma_{ij}.   
\end{align}
Here, $[\Ykr{k}{r}]_i$ denotes the $i$-th component of the modular form $\Ykr{k}{r}$. 
As shown in Eq.~\eqref{eq-YudeSU5}, $Y_d = Y_e^T$ up to the splitting at $\order{R_\Sigma}$ 
because
$\ol{d} \in \ol{5}$ and $Q \in 10$ while
$\ol{e} \in 10$ and $\ell \in \ol{5}$.
Note that the number of coefficients may change depending 
on the number of modular forms for a given modular weight.  

\subsection{Assigning modular weights} 

The minimal choice to obtain the realistic Yukawa matrices is 
\begin{align}
 k_{10} = (2,3), \quad k_{5}= (0,0,2),  
\end{align}
and we take $(s,b) = (1,0)$ for $\ol{5}_i = 1^s_b$, 
so that the Yukawa matrices are given by 
\begin{align}
 Y_u =&\ \frac{1}{\sqrt{6}}
\begin{pmatrix}
  \sqrt{6}\alpha_1 Y^{(4)}_{1^0_1}  &
  \sqrt{3}\alpha_{2}\left[\Ykr{5}{2^1_2}\right]_2 & 
- \sqrt{3}\alpha_{2}\left[\Ykr{5}{2^1_2}\right]_1  \\  
 \sqrt{3} \alpha_{2} \left[\Ykr{5}{2^1_2}\right]_2 & 
 -\sqrt{2} \alpha_3^i \left[\Ykr{6}{3^0,i}\right]_3 & 
  \alpha_3^i \left[\Ykr{6}{3^0,i}\right]_2  \\   
-\sqrt{3} \alpha_{2}\left[\Ykr{5}{2^1_2}\right]_1  & 
         \alpha_3^i\left[\Ykr{6}{3^0,i}\right]_2   & 
 \sqrt{2}\alpha_3^i \left[\Ykr{6}{3^0,i}\right]_1  \\ 
\end{pmatrix}, \\ 
 Y_d = &\ \frac{1}{\sqrt{2}} 
\begin{pmatrix}
 \sqrt{2} \beta_{11}^d \Ykr{2}{1^1_2} & 
 \beta_{12}^d  \left[\Ykr{3}{2^{0}_0}\right]_2 & 
-\beta_{12}^d  \left[\Ykr{3}{2^{0}_0}\right]_1  
 \\ 
\sqrt{2} \beta_{21}^d  \Ykr{2}{1^{1}_2} & 
 \beta_{22}^d  \left[\Ykr{3}{2^{0}_0}\right]_2 & 
-\beta_{22}^d  \left[\Ykr{3}{2^{0}_0}\right]_1  
 \\ 
0 & 
 \beta_{32}^d  \left[\Ykr{5}{2^{0}_0}\right]_2 & 
-\beta_{32}^d \left[\Ykr{5}{2^{0}_0}\right]_1 & 
 \\ 
\end{pmatrix}, 
\\ \notag 
 Y_e =&\  \frac{1}{\sqrt{2}} 
\begin{pmatrix}
 \sqrt{2} \beta_{11}^e \Ykr{2}{1^1_2} & 
 \sqrt{2} \beta_{21}^e  \Ykr{2}{1^{1}_2} & 
 0 \\ 
 \beta_{12}^e  \left[\Ykr{3}{2^{0}_0}\right]_2 & 
 \beta_{22}^e  \left[\Ykr{3}{2^{0}_0}\right]_2 & 
 \beta_{32}^e  \left[\Ykr{5}{2^{0}_0}\right]_2 \\
-\beta_{12}^e  \left[\Ykr{3}{2^{0}_0}\right]_1  &
-\beta_{22}^e  \left[\Ykr{3}{2^{0}_0}\right]_1  &
-\beta_{32}^e \left[\Ykr{5}{2^{0}_0}\right]_1 \\
\end{pmatrix}. 
\end{align}
There are two modular forms $\Ykr{6}{3^0,i}$, $i=1,2$, 
and there is no $1^1_2$ at $k=4$. 
The Yukawa matrices are rescaled by the canonical normalization of the kinetic terms  
as in Eq.~\eqref{eq-Kahler}.  
Since the top Yukawa coupling is $\order{1}$, 
we fix the overall factor of the Yukawa couplings, 
so that the factor $(\sqrt{2\mathrm{Im}\tau})^6$ is compensated. 
Note that the normalization of the modular form is not fixed from the bottom-up approach~\footnote{
We also assume that this normalization factor is universal 
for all of the matter fields $\ol{5}_i$ and $10_i$, 
and hence do not change the flavor structure. 
}. 
We absorb this effect into the coefficients, rather than the modular forms, by defining 
\begin{align}
    \hat{\alpha}_{i} := (2\mathrm{Im\tau})^3\alpha_i, 
    \quad 
    \hat{\beta}_{ij} :=  (2\mathrm{Im\tau})^3\beta_{ij}, 
    \quad 
    \hat{\gamma}_{ij} := (2\mathrm{Im\tau})^3\gamma_{ij}. 
\end{align}

With the assignments of the modular weights, the texture of the Yukawa matrices are given by 
\begin{align}
Y_u \sim 
\begin{pmatrix}
    \eta \eps^2 & \eta^{1/2} \eps^3 & \eta^{1/2} \eps \\ 
    \eta^{1/2} \eps^3 & \eps^4 & \eps^2 \\ 
    \eta^{1/2} \eps & \eps^2 & 1 
\end{pmatrix}, 
\quad 
Y_d \sim Y_e^T \sim
\begin{pmatrix}
    \eta^2 \eps & \eta^{3/2} \eps^2 & \eta^{3/2} \\ 
    \eta^2 \eps & \eta^{3/2} \eps^2 & \eta^{3/2} \\ 
            0 & \eta^{1/2} \eps^2 & \eta^{1/2} 
\end{pmatrix}, 
\end{align}
where $\eta := 1/(2\mathrm{Im}\tau)$ is the factor from the canonical normalization. 
The hierarchical structures of the quark masses and the CKM matrix is given by 
\begin{align}
 (m_u, m_c, m_d, m_s, m_b)/m_t \sim&\ 
\left(\eps^4, \eta \eps^2, \eta^{3/2}\eps^2, \eta^2\eps, \eta^{1/2} \right) \\ \notag 
\sim&\ 
\left( 2\times 10^{-5}, 7\times 10^{-4}, 3\times10^{-4}, 2\times10^{-3}, 0.4 \right) 
, \\ \notag 
(s^Q_{12}, s^Q_{23}, s^Q_{13}) \sim&\  
\left(\eps/\eta^{1/2}, \eta^{1/2}\eps, \eps^2\right) \sim 
\left(0.2, 0.03, 0.005\right), 
\end{align}
where $\eps = 0.067$ and $\eta = 0.16$   
are used for the numerical estimations. 
Here, $s_{ij}^Q$ is the mixing angles in the standard parametrization of the CKM matrix. 
The mass hierarchies of the charged leptons are the same as those for the down-type quarks. 
These values well fit to the data shown in Table~\ref{tab-obsCF} at the benchmark points 
discussed in the next section, 
except for $y_u$ and $y_e$ which are about 10 larger than the experimental values.  
These differences will be explained by the numerical factors in the modular forms 
and the $\order{1}$ coefficients. 
It is interesting that the CKM angles in our models fit to the data 
after taking account the powers of $\eta$ from the canonical normalization, 
so that $s_{23}^Q/s_{12}^Q \sim \eta \sim 0.2$.    

\subsection{Benchmark points}

\begin{table} 
\centering 
\caption{\label{tab-obsCF}
The values of the Yukawa couplings of the SM fermions  
and the mixing angles in the CKM matrix at the BP1 and BP2 
in the left and right panels, respectively. 
The second column is the value at the benchmark points, 
and the third (fourth) column is the experimental value (and its error). 
The experimental values are given at the GUT scale calculated 
in Ref.~\cite{Antusch:2013jca}. 
} 
\begin{minipage}[t]{0.48\hsize}
\centering
\begin{tabular}[t]{c|ccc} \hline 
obs. & value & center & error \\  \hline \hline 
$y_u$$/10^{-6}$ & 2.73 & 2.74 & 0.85 \\  
$y_c$$/10^{-3}$ & 1.391 & 1.421 & 0.050 \\  
$y_t$ & 0.5041 & 0.5033 & 0.0050 \\  \hline 
$y_d$$/10^{-5}$ & 5.94 & 5.90 & 0.65 \\  
$y_s$$/10^{-3}$ & 1.145 & 1.167 & 0.063 \\  
$y_b$$/10^{-2}$ & 5.375 & 5.388 & 0.054 \\  \hline 
$y_e$$/10^{-5}$ & 2.389 & 2.389 & 0.014 \\  
$y_\mu$$/10^{-3}$ & 5.049 & 5.043 & 0.030 \\  
$y_\tau$$/10^{-2}$ & 8.611 & 8.622 & 0.086 \\  \hline 
$s_{12}^Q$ & 0.22540 & 0.22541 & 0.00072 \\  
$s_{23}^Q$$/10^{-2}$ & 4.770 & 4.769 & 0.076 \\  
$s_{13}^Q$$/10^{-3}$ & 4.13 & 4.15 & 0.15 \\  
$\delta_{\mathrm{CKM}}$ & 1.2270 & 1.2080 & 0.0540 \\  \hline 
\end{tabular}
\end{minipage}
\begin{minipage}[t]{0.48\hsize}
\centering 
\begin{tabular}[t]{c|ccc} \hline 
obs. & value & center & error \\  \hline \hline 
$y_u$$/10^{-6}$ & 2.89 & 2.74 & 0.85 \\  
$y_c$$/10^{-3}$ & 1.450 & 1.421 & 0.050 \\  
$y_t$ & 0.5008 & 0.5029 & 0.0050 \\  \hline 
$y_d$$/10^{-5}$ & 7.12 & 5.87 & 0.65 \\  
$y_s$$/10^{-3}$ & 1.053 & 1.161 & 0.063 \\  
$y_b$$/10^{-2}$ & 3.977 & 3.983 & 0.040 \\  \hline 
$y_e$$/10^{-5}$ & 2.377 & 2.378 & 0.014 \\  
$y_\mu$$/10^{-3}$ & 5.000 & 5.019 & 0.030 \\  
$y_\tau$$/10^{-2}$ & 8.674 & 8.582 & 0.086 \\  \hline 
$s_{12}^Q$ & 0.22558 & 0.22541 & 0.00072 \\  
$s_{23}^Q$$/10^{-2}$ & 6.47 & 6.42 & 0.10 \\  
$s_{13}^Q$$/10^{-3}$ & 5.61 & 5.58 & 0.20 \\  
$\delta_{\mathrm{CKM}}$ & 1.2335 & 1.2080 & 0.0540 \\  \hline 
\end{tabular} 
\end{minipage}
\end{table}

We find the numerical values of the parameters by numerical optimization. 
We restrict the parameter space to be 
\begin{align}
 \tan\ol{\beta}\in (5, 60), 
\quad 
 \ol{\eta}_b \in (-0.6, 0.6), 
\quad 
\abs{\hat{\gamma}_{ij}} \in (-1,1).   
\end{align}  
Here, $\tan\ol{\beta}$ includes the threshold correction to the tau lepton, 
and $\ol{\eta}_b$ is that for the bottom quark as defined in Ref.~\cite{Antusch:2013jca}.
In this work, we treat $\ol{\eta}_b$ as a parameter which will be determined from the soft SUSY breaking parameters, see for threshold corrections Refs.~\cite{Hall:1993gn,Hempfling:1993kv}.
The threshold corrections to the light flavors, $\ol{\eta}_q$ and $\ol{\eta}_\ell$, 
are assumed to be zero for simplicity.  
We take $R_\Sigma = 0.1$ and restrict $\abs{\hat{\gamma}_{ij}} < 1$, 
so that the contribution from the dimension-5 operator is sub-dominant.

We found the following two benchmark points. 
At the first benchmark point (BP1), the inputs are given by 
$\tan\overline{\beta} = 11.4643$,  $\overline{\eta}_b = 0.187818$ 
$\tau = 0.0592+3.1033i$, 
\begin{align} 
\begin{pmatrix} 
\hat{\alpha}_{1} \\ \hat{\alpha}_2 \\ \hat{\alpha}^1_3 \\ \hat{\alpha}^2_3 
\end{pmatrix} 
=&\ 
\begin{pmatrix} 
2.1054 \\ -1.9005 \\ -1.5069 \\ 1.7427 
\end{pmatrix} 
, \quad 
\begin{pmatrix} 
\hat{\beta}_{11} \\ \hat{\beta}_{12} \\ \hat{\beta}_{21} \\ \hat{\beta}_{22} \\ \hat{\beta}_{32} 
\end{pmatrix} 
= 
\begin{pmatrix} 
2.2077e^{0.1766i} \\ 1.8263 \\ -0.6111 \\ 0.2838 \\ 0.1922 
\end{pmatrix} 
, \quad 
\begin{pmatrix} 
\hat{\gamma}_{11} \\ \hat{\gamma}_{12} \\ \hat{\gamma}_{21} \\ \hat{\gamma}_{22} \\ \hat{\gamma}_{32} 
\end{pmatrix} 
= 
\begin{pmatrix} 
0.2568 \\ 0.8436 \\ -0.9999 \\ 0.2433 \\ -0.8884 
\end{pmatrix},  
\end{align} 
and at the second point (BP2), 
$\tan\overline{\beta} = 11.4303$,  $\overline{\eta}_b = 0.598145$,  
$\tau = 0.0661+3.0791i$, 
\begin{align} 
\begin{pmatrix} 
\hat{\alpha}_{1} \\ \hat{\alpha}_2 \\ \hat{\alpha}^1_3 \\ \hat{\alpha}^2_3 
\end{pmatrix} 
=&\ 
\begin{pmatrix} 
1.5437 \\ -1.4035 \\ -1.4993 \\ 1.3019 
\end{pmatrix} 
, \quad 
\begin{pmatrix} 
\hat{\beta}_{11} \\ \hat{\beta}_{12} \\ \hat{\beta}_{21} \\ \hat{\beta}_{22} \\ \hat{\beta}_{32} 
\end{pmatrix} 
= 
\begin{pmatrix} 
1.7296e^{0.2134i} \\ 1.2887 \\ -0.5323 \\ 0.2484 \\ 0.2092 
\end{pmatrix} 
, \quad 
\begin{pmatrix} 
\hat{\gamma}_{11} \\ \hat{\gamma}_{12} \\ \hat{\gamma}_{21} \\ \hat{\gamma}_{22} \\ \hat{\gamma}_{32} 
\end{pmatrix} 
= 
\begin{pmatrix} 
0.9997 \\ 0.7726 \\ -0.9332 \\ 0.3163 \\ -0.9619 
\end{pmatrix}. 
\end{align} 
These points realize the quark and charged lepton masses, 
and the CKM angles as shown in Table~\ref{tab-obsCF}. 
At both points, all of the observables are within $2\sigma$ range, 
and the largest discrepancy is $0.61\sigma$ ($1.93\sigma$) at the BP1 (BP2) 
for the charm (down) quark mass.   
The values of the parameters are similar at both points, 
but $\ol{\eta}_b$ is relatively small (large) at the BP1 (BP2). 
The absolute values of the coefficients 
are  in the range of $[0.19, 2,2]$ and $[0.20, 1.7]$ at BP1 and BP2, respectively. 
Thus, the $\order{1}$ coefficients can explain the hierarchies with the good accuracy.

\section{Neutrino sector} 
\label{sec-neut} 

We shall consider the neutrino sector in this section.  
We assume that the neutrinos are Majorana, 
so the neutrino masses are given by the Weinberg operator $W\ni (\ell H_u)^2$ 
at low-energies. 
With our choice of the representations and the weights of $\ell \in \ol{5}$, 
the masses of the neutrinos and the mixing angles in the PMNS matrix 
are predicted to be 
\begin{align}
 (m_{\nu_1}, m_{\nu_2}, m_{\nu_3}) \sim (\eta^2, \eta^2, 1) \sim (0.03, 0.03, 1),  
\quad 
 (s_{12}, s_{23}, s_{13}) \sim (1, \eta, \eta) \sim (1, 0.2, 0.2),  
\end{align}
and thus the neutrino observables will have the texture 
\begin{align}
R^{21}_{32} := \frac{m_{\nu_2}^2-m_{\nu_1}^2}{m_{\nu_3}^2-m_{\nu_1}^2} 
                \sim \eta^4 \sim 0.0007, 
\quad 
 (s_{12}^2, s_{23}^2, s_{13}^2) \sim (1, \eta^2, \eta^2) \sim (1, 0.03, 0.03),  
\end{align}
independently to the UV completion of the Weinberg operator. 
Hence, the angle $s_{13}^2$ is naturally explained, 
while the ratio of the mass squared differences $R^{21}_{32}$ 
and $s_{23}^2$ are predicted to be about an order of magnitude smaller
than the observed values.   
We will discuss how these discrepancies are explained in 
an explicit model based on the type-I seesaw mechanism.

For illustration, 
we assume the type-I seesaw mechanism to realize the tiny neutrino masses 
by introducing the three generations of singlets $N_i$.  
The superpotential is given by 
\begin{align}
 W = \frac{1}{2} N^T M_N N + N^T Y_n \ol{5}_A H^A
  \ni   \frac{1}{2} N^T M_N N + N^T Y_n \ell H_u. 
\end{align}
We choose the representations and modular weights of the right-handed neutrinos as 
\begin{align}
 N = 1^1_0 \oplus 2_0 =: N_1 \oplus N_2, \quad k_N = (0,2). 
\end{align}
The modular invariant superpotential is given by
\begin{align}
 W =&\ \frac{M_0}{2} \left[ 
A_1 N_1 N_1 + 2 A_2\left( \Ykr{2}{2_0} N_1 N_2 \right)_1 + 
A_3 \Ykr{4}{1^0_0} \left( N_2 N_2\right)_1 
+ A_4 \left(\Ykr{4}{2_0} \left( N_2 N_2\right)_{2_0}\right)_1
  \right]    \\ \notag 
  &\ + \sum_{i=1,2} \left[  B_{1i} \left(N_1\ol{5}_i\right)_1  
     + B_{2i} \left(\Ykr{2}{2_0} N_2 \ol{5}_i \right)_1  \right] H
     + B_{23} \left(\Ykr{4}{2_0} N_2 \ol{5}_3 \right)_1 H,  
\end{align}
where $A_i$ and $B_{ij}$ are the $\order{1}$ coefficients, 
and $M_0$ is the overall scale of the Majorana mass term. 
$(N_2 N_2)_{2_0}$ takes the combination of the representation $2_0$. 
Here, we take $\Ykr{0}{1^0_0} = 1$. There is no $1^0_0$ at $k=2$. 
The Majorana mass matrix $M_N$ and the neutrino Yukawa matrix $Y_n$ are given by 
\begin{align}
 M_N =&\ \frac{M_0}{{2}} 
\begin{pmatrix}
2 A_1  & 
 -\sqrt{2} A_2 \left[\Ykr{2}{2_0}\right]_2 & 
  \sqrt{2} A_2 \left[\Ykr{2}{2_0}\right]_1 
\\ 
-\sqrt{2} A_2 \left[\Ykr{2}{2_0}\right]_2 & 
\sqrt{2} A_3 \Ykr{4}{1^0_0} - A_4 \left[\Ykr{4}{2_0}\right]_1 & 
A_4 \left[\Ykr{4}{2_0}\right]_2 &
\\
\sqrt{2} A_2 \left[\Ykr{2}{2_0}\right]_1  & 
A_4 \left[\Ykr{4}{2_0}\right]_2 &
\sqrt{2} A_3 \Ykr{4}{1^0_0} + A_4 \left[\Ykr{4}{2_0}\right]_1 \\  
\end{pmatrix}, 
\\ \notag 
Y_n =&\ \frac{1}{\sqrt{2}} 
\begin{pmatrix}
\sqrt{2}  B_{11}  & 
\sqrt{2}  B_{12}  & 
0 
\\
- B_{21} \left[\Ykr{2}{2_0}\right]_2 & 
- B_{22} \left[\Ykr{2}{2_0}\right]_2 & 
- B_{23} \left[\Ykr{4}{2_0}\right]_2   
\\ 
  B_{21} \left[\Ykr{2}{2_0}\right]_1 & 
  B_{22} \left[\Ykr{2}{2_0}\right]_1 & 
  B_{23} \left[\Ykr{4}{2_0}\right]_1  
\end{pmatrix}. 
\end{align}
The neutrino mass matrix is given by 
\begin{align}
 M_\nu = v_u^2 \hat{Y}_n^T \hat{M}_N^{-1} \hat{Y}_n,  
\end{align}
where $\hat{Y}_n$ and $\hat{M}_N$ are after the canonical normalization. 
For the neutrino observables, the elements suppressed by $\eps \sim \order{0.05}$  
are irrelevant, 
and hence the matrices are approximately given by 
\begin{align}
\frac{\hat{M}_N}{M_0 (2\mathrm{Im}\tau)^2}  \sim&\ 
\begin{pmatrix}
 A_1 \eta^2 & 0 & -\dfrac{A_2}{2} \eta \\ 
 0 & \dfrac{A_3}{\sqrt{6}}  + \dfrac{A_4}{4\sqrt{2}} & 0 \\
-\dfrac{A_2}{2} \eta & 0 & \dfrac{A_3}{\sqrt{6}}  - \dfrac{A_4}{4\sqrt{2}}
\end{pmatrix}, 
\  
\frac{\hat{Y}_n}{(2\mathrm{Im}\tau)^2} \sim 
\begin{pmatrix}
 B_{11} \eta^2 & B_{12} \eta^2& 0 \\ 
 0       & 0 & 0  \\
-\dfrac{B_{21}}{2} \eta & - \dfrac{B_{22}}{2} \eta & -\dfrac{B_{23}}{4}    
\end{pmatrix}.  
\end{align}
Since the neutrino mass matrix $M_\nu$ is rank-2, 
the lightest neutrino mass appears only at $\order{\eps^3}$~\footnote{
If $N$ is assigned to be a triplet, 
the neutrino mass will be rank-1 for $\eps \to 0$, 
and thus the observed pattern is more difficult to be realized. 
}.    
The ratio of the heavier two neutrinos are given by 
\begin{align}
\label{eq-R23}
\sqrt{R^{21}_{32}} \simeq
 \frac{m_{\nu_2}}{m_{\nu_3}} 
\sim 
\frac{16(B_{11}^2+B_{22}^2)}{B_{23}^2A_1} 
\left(
\frac{A_3}{\sqrt{6}} - \frac{A_4}{4\sqrt{2}} - \frac{A_2^2}{4A_1}
\right) \eta^2,   
\end{align}
at the leading order in $\eta$. 
Because of the hierarchy in the neutrino masses, 
the model predicts the normal ordering~\cite{Workman:2022ynf}. 
The ratio is enhanced by $(B_{11}^2+B_{22}^2)/(B_{23}/4)^2 \sim \order{10}$  
coming from the Dirac Yukawa matrix $Y_n$ for all $A,~B={\cal O}(1)$,  
and thus the ratio of the mass squared difference $R^{12}_{23}$ 
is enhanced by $\order{100}$, consistent with the observed value $\sim 0.03$. 
While the mild discrepancy of $s_{23}^2$ will be explained 
simply by $\order{5}$ ratios of the coefficients.

\begin{table} 
\centering 
\caption{\label{tab-Nuobs}
The values of the neutrino observables at the benchmark points. 
 } 
\begin{minipage}[t]{0.48\hsize}
\begin{tabular}[t]{c|ccc} \hline 
obs. & value & center & error \\  \hline \hline 
$R^{21}_{32}$$/10^{-2}$ & 3.070 & 3.070 & 0.084 \\  
$s_{12}^2$ & 0.307 & 0.307 & 0.013 \\  
$s_{23}^2$ & 0.544 & 0.546 & 0.021 \\  
$s_{13}^2$$/10^{-2}$ & 2.200 & 2.200 & 0.070 \\  
$\delta_{\mathrm{PMNS}}$$/10^{0}$ & -3.02 & -2.01 & 0.63 \\  \hline 
\end{tabular}  
\end{minipage}
\begin{minipage}[t]{0.48\hsize}
\begin{tabular}[t]{c|ccc} \hline 
obs. & value & center & error \\  \hline \hline 
$R^{21}_{32}$$/10^{-2}$ & 3.069 & 3.070 & 0.084 \\  
$s_{12}^2$ & 0.307 & 0.307 & 0.013 \\  
$s_{23}^2$ & 0.546 & 0.546 & 0.021 \\  
$s_{13}^2$$/10^{-2}$ & 2.200 & 2.200 & 0.070 \\  
$\delta_{\mathrm{PMNS}}$$/10^{0}$ & -3.00 & -2.01 & 0.63 \\  \hline 
\end{tabular} 
\end{minipage}
\end{table}

We find the values to explain the neutrino observables at the benchmark points. 
At the BP1, the fitted values are 
$M_0 = 1.2819\times 10^{16}$ GeV, 
\begin{align} 
\begin{pmatrix} 
A_1 \\ A_2 \\ A_3 \\ A_4 
\end{pmatrix} 
=&\ 
\begin{pmatrix} 
1.8038 \\ -1.3098 \\ 1.0235 \\ -4.0367 
\end{pmatrix} 
, \quad 
\begin{pmatrix} 
B_{11} \\ B_{12} \\ B_{21} \\ B_{22} \\ B_{23} 
\end{pmatrix} 
= 
\begin{pmatrix} 
-2.4174 \\ 1.0236 \\ 4.0214 \\ 1.0236 \\ 2.6574 
\end{pmatrix}, 
\end{align} 
and at BP2, 
$M_0 = 1.7775\times 10^{15}$ GeV, 
\begin{align} 
\begin{pmatrix} 
A_1 \\ A_2 \\ A_3 \\ A_4 
\end{pmatrix} 
=&\ 
\begin{pmatrix} 
1.8984 \\ -0.6240 \\ -0.4992 \\ 1.1528 
\end{pmatrix} 
, \quad 
\begin{pmatrix} 
B_{11} \\ B_{12} \\ B_{21} \\ B_{22} \\ B_{23} 
\end{pmatrix} 
= 
\begin{pmatrix} 
0.8983 \\ -0.7126 \\ 1.8984 \\ 0.6480 \\ -0.3344 
\end{pmatrix}.  
\end{align} 
At these points, the CP phase in the PMNS matrix 
is originated from that in the charged lepton Yukawa matrix,  
which is common to the CKM matrix.  
We see that the values of the coefficients are close to $1$ 
and the ratio of the coefficients are at most $3.9$ ($5.7$) at the BP1 (BP2).

\section{Summary}
\label{sec-summary}

In this work, we build a model to realize the fermion hierarchy   
in $SU(5)$ GUT utilizing the modular $\Gamma_6^\prime$ symmetry.  
The residual symmetry associated with the generator $T$, namely $\intZ_6^T$ symmetry, 
controls the powers of the small parameter $e^{-2\pi\mathrm{Im}\tau/6}$.  
We argue that the $\Gamma_6^\prime$ is the minimal possibility 
to realize the hierarchical masses of the up-type quarks, 
because the Yukawa matrix for the up-type quarks is symmetric in the $SU(5)$ GUT. 
We assign the representations of $10$ as $1^0_1\oplus 2^{0}_2$ 
and $\ol{5}$ as $1^s_b\oplus 1^s_b \oplus 1^s_b$.  
We have to consider the double covering $\Gamma_6^\prime$ to have the representations $2^s_b$. 
The CKM matrix is hierarchical 
because the $T$ charges of $Q \in 10$ are different, 
while the PMNS matrix is not hierarchical 
because the those of $\ell \in \ol{5}$ are the same. 
The representations of $r$ and $(s,b)$ are chosen such that 
the realistic Yukawa matrices are realized for the modular weights smaller than $6$.  
We assigned the modular weights, 
so that certain modular forms exists and the Yukawa matrices are rank 3 
up to $\order{\eps^6}$. 
In the model, we show that the $\order{1}$ coefficients without hierarchies 
explain the observed quark and lepton masses and the CKM elements.

We assume the type-I seesaw mechanism to explain the tiny neutrino masses. 
The singlet right-handed neutrinos are assigned to be $1^1_0 \oplus 2_0$ 
and their modular weights are chosen to be $0\oplus 2$.  
With these assignments, 
one of the three neutrinos is suppressed by $\order{\eps^3}$ compared 
with the other two heavier neutrinos. 
The mass ratio of the heavier two neutrinos is not suppressed by $\eps$, 
but is suppressed by $(\mathrm{Im}\tau/2)^2 \sim \order{10}$ 
due to the canonical normalization, see Eq.~\eqref{eq-R23}. 
We showed that the $\order{1}$ coefficients can explain the neutrino 
observables in the $SU(5)$ GUT model.

\section*{Acknowledgment} 
The work of J.K. is supported in part by the Institute for Basic Science (IBS-R018-D1). 
This work is supported in part by he Grant-in-Aid for Scientific Research from the
Ministry of Education, Science, Sports and Culture (MEXT), Japan 
No.\ JP22K03601~(T.H.) and JP23K03375~(T.K.). 
The work of Y.A. is supported by JSPS Overseas Research Fellowships.

\appendix

\section{$\Gamma_6^\prime$ modular symmetry} 
\label{app-mod6}   

\subsection{Group theory of $\Gamma_6^\prime$} 

The algebra of $\Gamma_6^\prime$ is given by~\cite{Li:2021buv}
\begin{align}
 S^2 = R, \quad TR=RT, \quad R^2 = T^6 = (ST)^3 =  ST^2ST^3 ST^4ST^3=1. 
\end{align}
That for $\Gamma_6 = \Gamma_6^\prime/\intZ^R_2$ is given by taking $R=1$. 
The irreducible representations of $\Gamma_6^\prime$ are given by 
\begin{align}
 1^s_b, 2_b, 3^s, 6, 
\quad\mathrm{and}\quad
 2^s_b, 4_b, 
\end{align}
where $s=0,1$ and $b=0,1,2$ 
correspond to the $S_3$ and $A_4^\prime$ indices respectively, 
since $\Gamma_6^\prime$ is isomorphic to $S_3\times A_4^\prime$.  
The latter two representations exist only in $\Gamma_6^\prime$.  
The representation matrices are given by 
\begin{align}
 \rho_S(1^{s}_b) = (-1)^s, 
\quad  
 \rho_T(1^{s}_b) = (-1)^s w^b,  
\end{align}
for the singlet $1^s_b$, 
\begin{align}
 \rho_S(2_b) = \frac{1}{2} 
\begin{pmatrix}
- 1 & \sqrt{3} \\ \sqrt{3} & 1 
\end{pmatrix}, 
\quad 
 \rho_T(2_b) = w^b 
\begin{pmatrix}
1 & 0 \\ 0 & -1 
\end{pmatrix}, 
\end{align}
for the doublet $2_b$, 
\begin{align}
\rho_S(3^s) = (-1)^s \frac{1}{3}
\begin{pmatrix}
 -1 & 2 & 2 \\ 2 & -1 & 2 \\ 2 & 2 & -1 
\end{pmatrix}, 
\quad 
\rho_T(3^s) = (-1)^s 
\begin{pmatrix}
 1 & 0 & 0 \\ 0 & w & 0 \\ 0 & 0 & w^2 
\end{pmatrix},  
\end{align}
for the triplet $3^s$ 
and 
\begin{align}
 \rho_S(6) = \frac{1}{2} 
\begin{pmatrix}
- \rho_S(3^0)         &  \sqrt{3} \rho_S(3^0) \\ 
 \sqrt{3}\rho_S(3^0) &   \rho_S(3^0) \\ 
\end{pmatrix}, 
\quad 
 \rho_T(6) = 
\mathrm{diag}\left(1,w,w^2,-1, -w,-w^2 \right),  
\end{align}
for the sextet $6$, where $w := e^{2\pi i/3}$. 
The representation matrices of $2^s_b$ and $4_b$ are respectively given by 
\begin{align}
 \rho_S(2^s_b) = (-1)^s 
\frac{i}{\sqrt{3}}
\begin{pmatrix}
 1 & \sqrt{2} \\ \sqrt{2} & -1 
\end{pmatrix}, 
\quad 
 \rho_T(2^s_b) = (-1)^s w^b 
\begin{pmatrix}
 1 & 0 \\ 0 & w 
\end{pmatrix},   
\end{align}
and 
\begin{align}
 \rho_S(4_b) = 
\frac{1}{2}
\begin{pmatrix}
 -\rho_S(2^0_b) & \sqrt{3}\rho_S(2^0_b) \\ 
 \sqrt{3}\rho_S(2^0_b) & \rho_S(2^0_b) 
\end{pmatrix}, 
\quad 
 \rho_T(4_b) = 
\begin{pmatrix}
 \rho_T(2^0_b) & 0 \\ 0 & -\rho_T(2^0_b)  
\end{pmatrix}.   
\end{align} 
Note that the definition of $2^s_b$ is different from that in Ref.~\cite{Li:2021buv}, 
where $\rho_T(2^s_b) \propto w^{b+1}$.

The direct product of the singlets is given by 
\begin{align}
 1^s_b \otimes 1^t_c = 1^{s+t}_{b+c}. 
\end{align}
Here and hereafter, $s+t$ ($b+c$) should be understood as modulo $2$ ($3$). 
The products involving the singlet are given by 
\begin{align}
1^s_b(\alpha) \otimes 2_c (\beta) 
   = 
 P_2^s  
 \begin{pmatrix}
\alpha  \beta_1 \\ \alpha \beta_2
 \end{pmatrix}_{2_{b+c}}, 
\ 
1^{s}_b (\alpha) \otimes 3^s(\beta)
=  P_3^b    
   \begin{pmatrix}
  \alpha\beta_1 \\ \alpha\beta_2 \\ \alpha\beta_3
 \end{pmatrix}_{3^{r+s}}, 
\ 
 1^s_b (\alpha) \otimes 6(\beta)
=  P_{sb}
\begin{pmatrix}
 \alpha\beta_1 \\ \alpha\beta_2 \\ \alpha\beta_3 \\ 
 \alpha\beta_4 \\ \alpha\beta_5 \\  \alpha\beta_6 
\end{pmatrix}_6,  
\notag 
\end{align}
and 
\begin{align}
 1^s_b \otimes 2^t_c = 2^{s+t}_{b+c}, 
\quad 
 1^s_b(\alpha) \otimes 4_c(\beta) = P_4^s 
\begin{pmatrix}
 \alpha \beta_1 \\ \alpha \beta_2 \\ 
 \alpha \beta_3 \\ \alpha \beta_4 \\
\end{pmatrix}_{4_{b+c}},  
\end{align}
where 
\begin{align}
 P_2 := 
\begin{pmatrix}
 0 & 1 \\ -1 & 0 
\end{pmatrix},  
\quad 
P_3 := 
\begin{pmatrix}
 0 & 0 & 1 \\ 1 & 0 & 0  \\ 0 & 1 & 0 
\end{pmatrix}, 
\quad 
 P_{sb} := 
\begin{pmatrix}
 0 & \id{3} \\ -\id{3} & 0 
\end{pmatrix}^s 
\begin{pmatrix}
 P_3 & 0 \\ 0 & P_3 
\end{pmatrix}^b, 
\quad 
  P_4 := 
\begin{pmatrix}
 0 & \id{2} \\ -\id{2} & 0
\end{pmatrix}. 
\end{align}

The products of the doublets are given by 
\begin{align}
 \begin{pmatrix}
  \alpha_1 \\ \alpha_2 
 \end{pmatrix}_{2_b}
\otimes 
 \begin{pmatrix}
  \beta_1 \\ \beta_2 
 \end{pmatrix}_{2_c} 
=&\ \frac{1}{\sqrt{2}}
\left(\alpha_1\beta_1+\alpha_2\beta_2\right)_{1^{0}_{b+c}}
\oplus \frac{1}{\sqrt{2}}
\left(\alpha_1\beta_2-\alpha_2\beta_1\right)_{1^{1}_{b+c}}
\oplus \frac{1}{\sqrt{2}}
\begin{pmatrix}
 \alpha_2\beta_2-\alpha_1\beta_1 \\ 
 \alpha_1\beta_2+\alpha_2\beta_1 
\end{pmatrix}_{2_{b+c}}, 
 \notag  \\ 
 \begin{pmatrix}
  \alpha_1 \\ \alpha_2
 \end{pmatrix}_{2^{s}_b}
\otimes 
 \begin{pmatrix}
  \beta_1 \\ \beta_2
 \end{pmatrix}_{2_c}
=&\ P_4^{s}  
 \begin{pmatrix}
  \alpha_1\beta_1 \\ \alpha_2 \beta_1 \\ \alpha_1 \beta_2 \\ \alpha_2\beta_2
 \end{pmatrix}_{4_{b+c}}, 
\\ \notag 
  \begin{pmatrix}
  \alpha_1 \\ \alpha_2
 \end{pmatrix}_{2^s_b}
\otimes 
 \begin{pmatrix}
  \beta_1 \\ \beta_2
 \end{pmatrix}_{2^t_c}
=&\ 
\frac{1}{\sqrt{2}}\left(\alpha_1\beta_2-\alpha_2\beta_1\right)_{1^{s+t}_{b+c+1}}
 \oplus P^{b+c}_3  \frac{1}{\sqrt{2}}
\begin{pmatrix}
 -\sqrt{2}\alpha_1\beta_1 \\ \alpha_1\beta_2 +\alpha_2\beta_1 \\ 
  \sqrt{2}\alpha_2\beta_2
\end{pmatrix}_{3^{s+t}}.  
\end{align}
The products of the doublet and triplet are 
\begin{align}
 \begin{pmatrix}
  \alpha_1 \\ \alpha_2 
 \end{pmatrix}_{2_b}
\otimes 
 \begin{pmatrix}
  \beta_1 \\ \beta_2  \\ \beta_3 
 \end{pmatrix}_{3^s}
 =&\  P_{sb}  
 \begin{pmatrix}
  \alpha_1 \beta_1 \\   \alpha_1 \beta_2 \\   \alpha_1 \beta_3 \\ 
  \alpha_2 \beta_1 \\   \alpha_2 \beta_2 \\   \alpha_2 \beta_3 \\ 
 \end{pmatrix}_6, 
\\ \notag 
 \begin{pmatrix}
  \alpha_1 \\ \alpha_2
 \end{pmatrix}_{2^{s}_b}
\otimes 
 \begin{pmatrix}
  \beta_1 \\ \beta_2 \\ \beta_3 
 \end{pmatrix}_{3^t} 
=&\  \frac{1}{\sqrt{3}}
 \begin{pmatrix}
  \alpha_1\beta_1+\sqrt{2} \alpha_2\beta_3 \\ 
  \sqrt{2} \alpha_1\beta_2-\alpha_2\beta_1  
 \end{pmatrix}_{2^{s+t}_{b}}
\oplus \frac{1}{\sqrt{3}}
 \begin{pmatrix}
  \alpha_1\beta_2+\sqrt{2} \alpha_2\beta_1 \\ 
  \sqrt{2} \alpha_1\beta_3-\alpha_2\beta_2  
 \end{pmatrix}_{2^{s+t}_{b+1}}
\\ \notag 
&\ \oplus \frac{1}{\sqrt{3}}
 \begin{pmatrix}
  \alpha_1\beta_3+\sqrt{2} \alpha_2\beta_2 \\ 
  \sqrt{2} \alpha_1\beta_1-\alpha_2\beta_3  
 \end{pmatrix}_{2^{s+t}_{b+2}},  
\end{align}
and the product of the triplets is given by 
\begin{align}
 \begin{pmatrix}
  \alpha_1 \\ \alpha_2  \\ \alpha_3  
 \end{pmatrix}_{3^s}  
\otimes 
 \begin{pmatrix}
  \beta_1 \\ \beta_2  \\ \beta_3 
 \end{pmatrix}_{3^t}
=&\ \frac{1}{\sqrt{3}}
\left(\alpha_1\beta_1+\alpha_2\beta_3+\alpha_3\beta_2\right)_{1^{s+t}_0} 
\oplus \frac{1}{\sqrt{3}}
\left(\alpha_3\beta_3+\alpha_1\beta_2+\alpha_2\beta_1\right)_{1^{s+t}_1} \\ \notag 
\oplus&\ \frac{1}{\sqrt{3}}
\left(\alpha_2\beta_2+\alpha_3\beta_1+\alpha_1\beta_3\right)_{1^{s+t}_2}   
\oplus
\frac{1}{\sqrt{6}}
\begin{pmatrix}
 2\alpha_1\beta_1-\alpha_2\beta_3-\alpha_3\beta_2 \\ 
 2\alpha_3\beta_3-\alpha_1\beta_2-\alpha_2\beta_1 \\ 
 2\alpha_2\beta_2-\alpha_3\beta_1-\alpha_1\beta_3 \\ 
\end{pmatrix}_{3^{s+b}_S} 
\\ \notag 
\oplus&\ 
\frac{1}{\sqrt{2}}
\begin{pmatrix}
 \alpha_2\beta_3-\alpha_3\beta_2 \\ 
 \alpha_1\beta_2-\alpha_2\beta_1 \\ 
 \alpha_3\beta_1-\alpha_1\beta_3 \\ 
\end{pmatrix}_{3^{s+b}_A}. 
\end{align}
The product rules involving the higher dimensional representations 
are shown in Ref.~\cite{Li:2021buv}.

\subsection{Modular forms} 
\begin{table}[t]
 \centering
\caption{\label{tab-rnum}
The representations of modular forms at $k\le 6$.}
\begin{tabular}[t]{c|c} \hline 
weight & representations \\ \hline\hline
$k=1$  & $2^0_0$, $4_2$ \\ 
$k=2$  & $1^1_2$, $2_0$, $3^0$, $6$ \\ 
$k=3$  & $2^0_0$, $2^0_2$, $2^1_2$, $4_0$, $4_1$, $4_2$ \\ 
$k=4$  & $1_0^0$, $1_1^0$, $2_0$, $2_2$, $3^0$, $3^1$, $6$, $6$ \\
$k=5$  & $2^0_0$, $2^0_1$, $2^0_2$, $2^1_1$, $2^1_2$, $4_0$, $4_0$, $4_1$, $4_2$, $4_2$ \\
$k=6$  & $1^0_0$, $1^1_0$, $1^1_2$, $2_0$, $2_1$, $2_2$, $3^0$, $3^0$, $3^1$, 
        $6$, $6$, $6$\\
\hline
\end{tabular} 
\end{table}

The modular forms at $k=1$ are given by~\cite{Li:2021buv} 
\begin{align}
Y^{(1)}_{2^0_0} =  
 \begin{pmatrix}
 Y_1 \\ Y_2
\end{pmatrix}_{2^0_0}
= 
\begin{pmatrix}
 3e_1 + e_2 \\ 3\sqrt{2}e_1
\end{pmatrix},
\quad 
Y^{(1)}_{4_2} =  
\begin{pmatrix}
  Y_3 \\ Y_4  \\ Y_5 \\ Y_6
\end{pmatrix}_{4_2} 
= 
\begin{pmatrix}
 3\sqrt{2} e_3 \\ 
 -3e_3 - e_5 \\ 
\sqrt{6} e_3  -\sqrt{6}e_6 \\ 
 -\sqrt{3}e_3 + 1/\sqrt{3} e_4
- 1/\sqrt{3} e_5+\sqrt{3}e_6
\end{pmatrix},   
\end{align}
where the functions are defined as 
\begin{align}
e_1(\tau):=&\ \frac{\eta(3\tau)^3}{\eta(\tau)}, 
\quad& 
e_2(\tau):=&\  \frac{\eta(\tau/3)^3}{\eta(\tau)},  
\quad& 
e_3(\tau):=&\ \frac{\eta(6\tau)^3}{\eta(2\tau)}, 
\\ \notag 
e_4(\tau):=&\ \frac{\eta(\tau/6)^3}{\eta(\tau/2)}, 
\quad& 
e_5(\tau):=&\ \frac{\eta(2\tau/3)^3}{\eta(2\tau)}, 
\quad& 
e_6(\tau):=&\ \frac{\eta(3\tau/2)^3}{\eta(\tau/2)}.  
\end{align}
The modular forms with higher weights can be constructed by taking the direct products.

The modular forms at weight $k=2$ are 
\begin{align}
Y^{(2)}_{1^1_2} =&\ \left(Y_3Y_6-Y_4Y_5\right)_{1^1_2},  
\quad 
Y^{(2)}_{2_0} = \frac{1}{\sqrt{2}}  
\begin{pmatrix}
 Y_1Y_4-Y_2Y_3 \\ Y_1Y_6-Y_2Y_5 
\end{pmatrix},
\\ \notag 
Y^{(2)}_{3^0}
=&\ 
\begin{pmatrix}
 -Y_1^2 \\ \sqrt{2}Y_1Y_2 \\ Y_2^2 
\end{pmatrix},   
\quad 
Y^{(2)}_{6} = \frac{1}{\sqrt{2}}  
\begin{pmatrix}
 Y_1Y_4+Y_2Y_3 \\
  \sqrt{2}Y_2Y_4 \\ 
 -\sqrt{2}Y_1Y_3 \\ 
 Y_1Y_6+Y_2Y_5 \\ 
  \sqrt{2} Y_2Y_6 \\ 
 -\sqrt{2} Y_1Y_5
\end{pmatrix}. 
\end{align}
At weight $k=3$, 
\begin{align}
Y^{(3)} _{2^0_0} =&\ \frac{1}{\sqrt{3}} 
\begin{pmatrix}
 -Y_1^3+\sqrt{2}Y_2^3 \\ 3Y_1^2Y_2 
\end{pmatrix}, 
\quad 
Y^{(3)} _{2^0_2} = \frac{1}{\sqrt{3}}  
\begin{pmatrix}
 3Y_1Y_2^2 \\ -\sqrt{2} Y_1^3-Y_2^3 
\end{pmatrix}, 
\quad 
Y^{(3)} _{2^1_2} = \left(Y_3Y_6-Y_4Y_5\right)
\begin{pmatrix}
Y_1 \\ Y_2 
\end{pmatrix},  
\notag  \\ 
Y^{(3)}_{4_0} =&\ \sqrt{\frac{2}{3}}
\begin{pmatrix}
 Y_1Y_2Y_3 - Y_1^2 Y_4 \\ 
 Y_2^2 Y_3 - Y_1Y_2Y_4 \\ 
 Y_1Y_2Y_5 - Y_1^2 Y_6 \\ 
 Y_2^2 Y_5 - Y_1Y_2Y_6 
\end{pmatrix}, 
\quad  
Y^{(3)}_{4_1} = \frac{1}{\sqrt{3}}
\begin{pmatrix}
 Y_2^2Y_3+2Y_1Y_2Y_4 \\ 
-Y_2^2Y_4-\sqrt{2}Y_1^2Y_3 \\ 
 Y_2^2Y_5+2Y_1Y_2Y_6 \\ 
-Y_2^2Y_6-\sqrt{2}Y_1^2Y_5  
\end{pmatrix}, 
\\ \notag 
Y^{(3)}_{4_2} =&\ \frac{1}{\sqrt{3}}
\begin{pmatrix}
-Y_1^2Y_3 + \sqrt{2}Y_2^2Y_4  \\ 
 Y_1^2Y_4 +        2Y_1Y_2Y_3 \\ 
-Y_1^2Y_5 + \sqrt{2}Y_2^2Y_6  \\  
 Y_1^2Y_6 +        2Y_1Y_2Y_5 
\end{pmatrix}, 
\end{align}
and at weight $k=4$, 
\begin{align}
 Y^{(4)}_{1_0^0} 
=&\ \frac{1}{\sqrt{3}} Y_1 \left(Y_1^3 + 2\sqrt{2} Y_2^3\right), 
\quad &
 Y^{(4)}_{1_1^0} 
=&\ \left(Y_3Y_6-Y_4Y_5\right)^2,   \\ \notag 
 Y^{(4)}_{2_0} =&\ \frac{1}{2\sqrt{2}} 
\begin{pmatrix}
 (Y_1Y_6-Y_2Y_5)^2-(Y_1Y_4-Y_2Y_3)^2 \\ 
 2(Y_1Y_4-Y_2Y_3)(Y_1Y_6-Y_2Y_5)
\end{pmatrix}, 
\quad& 
  Y^{(4)}_{2_2} =&\ \frac{1}{\sqrt{2}} 
\left(Y_3Y_6-Y_4Y_5\right)
\begin{pmatrix}
 Y_1Y_6-Y_2Y_5 \\ Y_2Y_3-Y_1Y_4
\end{pmatrix}, \\ \notag
Y^{(4)}_{3^0} =&\ \sqrt{\frac{2}{3}}
\begin{pmatrix}
 Y_1\left(Y_1^3-\sqrt{2} Y_2^3\right) \\ 
 Y_2\left(Y_2^3+\sqrt{2} Y_1^3\right) \\ 
 3Y_1^2Y_2^2 
\end{pmatrix}, 
\quad& 
Y^{(4)}_{3^1} =&\ \left(Y_3Y_6-Y_4Y_5\right)
\begin{pmatrix}
\sqrt{2}Y_1Y_2 \\ Y_2^2  \\  -Y_1^2  
\end{pmatrix}, 
\\ \notag 
Y^{1(4)}_{6,1}
=&\ \frac{1}{\sqrt{2}} 
\left(Y_3Y_6-Y_4Y_5\right)
\begin{pmatrix}
  \sqrt{2}Y_2Y_6 \\ -\sqrt{2}Y_1Y_5 \\ Y_1Y_6+Y_2Y_5 \\ 
- \sqrt{2}Y_2Y_4 \\ \sqrt{2}Y_1Y_3 \\ -Y_1Y_4-Y_2Y_3  
\end{pmatrix}, 
\quad& 
Y^{2(4)}_{6,2} =&\  \frac{1}{\sqrt{2}} 
\begin{pmatrix}
       -(Y_1Y_4-Y_2Y_3) Y_1^2 \\ 
\sqrt{2}(Y_1Y_4-Y_2Y_3) Y_1Y_2 \\ 
        (Y_1Y_4-Y_2Y_3) Y_2^2 \\ 
       -(Y_1Y_6-Y_2Y_5) Y_1^2 \\ 
\sqrt{2}(Y_1Y_6-Y_2Y_5) Y_1Y_2 \\ 
        (Y_1Y_6-Y_2Y_5) Y_2^2 \\  
\end{pmatrix}. 
\end{align}
At $k=5$, 
\begin{align}
\Ykr{5}{2^0_0} =&\ \frac{1}{\sqrt{3}}Y_1\left(Y_1^3+2\sqrt{2}Y_2^3\right) 
\begin{pmatrix}
 Y_1 \\ Y_2
\end{pmatrix}, 
\ & 
\Ykr{5}{2^0_1} =&\ \left(Y_3Y_6-Y_4Y_5\right)^2 
\begin{pmatrix}
 Y_1 \\ Y_2
\end{pmatrix},  
\\ \notag 
\Ykr{5}{2^0_2} =&\ \frac{1}{3}
\begin{pmatrix}
-5Y_1^3Y_2^2 - \sqrt{2}Y_2^5 \\ 
 5Y_1^2Y_2^3 - \sqrt{2} Y_1^5 
\end{pmatrix}, 
& 
\Ykr{5}{2^1_1} =&\ \frac{1}{\sqrt{3}}\left(Y_4Y_5-Y_3Y_6\right) 
\begin{pmatrix}
 -3Y_1Y_2^2 \\ \sqrt{2}Y_1^3 + Y_2^3 
\end{pmatrix},  
\\ \notag 
\Ykr{5}{2^1_2} =&\ \frac{1}{\sqrt{3}}\left(Y_3Y_6-Y_4Y_5\right) 
\begin{pmatrix}
 -Y_1^3+\sqrt{2}Y_2^3 \\ 3Y_1^2Y_2 
\end{pmatrix}, 
\quad& 
\quad 
\\ \notag 
\Ykr{5}{4_0,1} =&\ \frac{1}{\sqrt{3}}\left(Y_3Y_6-Y_4Y_5\right)
\begin{pmatrix}
 Y_2^2Y_5+2Y_1Y_2Y_6 \\ 
-Y_2^2Y_6-\sqrt{2}Y_1^2Y_5 \\ 
-Y_2^2Y_3-2Y_1Y_2Y_4 \\ 
Y_2^2Y_4+\sqrt{2}Y_1^2Y_3 
\end{pmatrix}, 
& 
\Ykr{5}{4_0,2} =&\ \left(Y_3Y_6-Y_4Y_5\right)^2 
\begin{pmatrix}
Y_3 \\ Y_4 \\ Y_5 \\ Y_6 
\end{pmatrix},  
\\ \notag 
\Ykr{5}{4_1} =&\ \frac{1}{\sqrt{3}}\left(Y_3Y_6-Y_4Y_5\right)  
\begin{pmatrix}
 -Y_1^2Y_5+\sqrt{2}Y_2^2Y_6 \\ 
 Y_1^2Y_6 + 2Y_1Y_2Y_5 \\ 
 Y_1^2Y_3-\sqrt{2}Y_2^2Y_4 \\ 
-Y_1^2Y_4-2Y_1Y_2Y_3 \\ 
\end{pmatrix}, 
\ &
\quad 
\\ \notag 
\Ykr{5}{4_2,1} =&\ \sqrt{\frac{2}{3}}\left(Y_3Y_6-Y_4Y_5\right)  
\begin{pmatrix}
Y_1Y_2Y_5-Y_1^2Y_6 \\ 
Y_2^2Y_5 - Y_1Y_2Y_6 \\ 
-Y_1Y_2Y_3+Y_1^2Y_4 \\ 
-Y_2^2Y_3   + Y_1Y_2Y_4 \\ 
\end{pmatrix}, 
\ & 
\Ykr{5}{4_2,2} =&\ \frac{1}{\sqrt{3}}Y_1\left(Y_1^3+2\sqrt{2}Y_2^3\right) 
\begin{pmatrix}
Y_3 \\ Y_4 \\ Y_5 \\ Y_6 
\end{pmatrix}.
\\ \notag 
\end{align}

At weight $k=6$, 
\begin{align}
 \Ykr{6}{1^0_0} =&\ 
\frac{1}{4\sqrt{2}} \left(Y_1Y_4-Y_2Y_3\right)  
                    \left\{ 3(Y_1Y_6-Y_2Y_5)^2-(Y_1Y_4-Y_2Y_3)^2\right\}, 
\\ \notag 
 \Ykr{6}{1^1_0} =&\ \left(Y_3Y_6-Y_4Y_5\right)^3, 
\quad 
 \Ykr{6}{1^1_2} = \frac{1}{\sqrt{3}}Y_1\left(Y_1^3+2\sqrt{2}Y_2^3\right)
                                         \left(Y_3Y_6-Y_4Y_5\right),
\\ \notag 
\Ykr{6}{2_0} =&\ \frac{1}{\sqrt{6}} Y_1\left(Y_1^3+2\sqrt{2}Y_2^3\right) 
\begin{pmatrix}
 Y_1Y_4-Y_2Y_3 \\ Y_1Y_6-Y_2Y_5 
\end{pmatrix}, 
\\ \notag 
 \Ykr{6}{2_1} =&\ \frac{1}{\sqrt{2}}\left(Y_3Y_6-Y_4Y_5\right)^2 
\begin{pmatrix}
Y_2Y_3-Y_1Y_4 \\ Y_2Y_5-Y_1Y_6 
\end{pmatrix}, 
\\ \notag 
\Ykr{6}{2_2} =&\  
 \frac{1}{2\sqrt{2}} \left(Y_3Y_6-Y_4Y_5\right) 
\begin{pmatrix}
 2(Y_1Y_4-Y_2Y_3)(Y_1Y_6-Y_2Y_5) \\ 
(Y_1Y_4-Y_2Y_3)^2-(Y_1Y_6-Y_2Y_5)^2 
\end{pmatrix}, 
\\ \notag 
 \Ykr{6}{3^0,1} =&\  
\frac{1}{\sqrt{3}} Y_1\left(Y_1^3+2\sqrt{2}Y_2^3\right) 
\begin{pmatrix}
 -Y_1^2 \\ \sqrt{2}Y_1Y_2 \\ Y_2^2 
\end{pmatrix}, 
\quad 
 \Ykr{6}{3^0,2} = 
\frac{1}{\sqrt{3}}
\begin{pmatrix}
 Y_2^6-2\sqrt{2}Y_1^3Y_2^3 \\ 
2\sqrt{2}Y_1^5Y_2 - Y_1^2Y_2^4 \\ 
-4Y_1^4Y_2^2 + \sqrt{2}Y_1Y_2^5 
\end{pmatrix}, 
\\ \notag 
\Ykr{6}{3^1} =&\ \sqrt{\frac{2}{3}} 
\left(Y_3Y_6-Y_4Y_5\right)
\begin{pmatrix}
Y_2(Y_2^3+\sqrt{2}Y_1^3) \\ 3Y_1^2Y_2^2  \\  Y_1(Y_1^3-\sqrt{2}Y_2^3)  
\end{pmatrix}, \\ 
\Ykr{6}{6,1} =& \Ykr{2}{1^1_2}\otimes \Ykr{4}{6,1}, 
\quad  
\Ykr{6}{6,2} = \Ykr{2}{1^1_2}\otimes \Ykr{4}{6,2}, 
\quad 
\Ykr{6}{6,3} = \Ykr{2}{2_0}\otimes \Ykr{4}{3^0}.  
\end{align}
The representations existing for $k\le 6$ are summarized in Table~\ref{tab-rnum}.

At $\tau \sim i\infty$, it is convenient to explained the modular forms 
by $q := e^{2\pi i \tau}$ whose absolute values is small.  
The $q$-expansions of the functions $e_i$ are given by 
\begin{align}
 e_1 =&\ q^{1/3} + q^{4/3}, \quad&  
 e_2 =&\ 1-3q^{1/3} + 6q - 3q^{4/3}, \\ \notag 
 e_3 =&\ q^{2/3}, \quad & 
 e_4 =&\ 1-3q^{1/6}+6q^{1/2}-3q^{2/3} -6q^{7/6} + 6q^{3/2}, \\ \notag 
 e_5 =&\ 1-3q^{2/3}, \quad& 
 e_6 =&\ q^{1/6}+q^{2/3} + 2q^{7/6}, 
\end{align}
where $\order{q^2}$ is neglected. 
Defining 
\begin{align}
\label{eq-epsdef}
 \eps := \sqrt{3}q^{1/6}, 
\end{align}
the modular forms up to $\order{\eps^5}$ are given by 
\begin{align}
\Ykr{1}{2^0_0} = 
\begin{pmatrix}
 Y_1 \\ Y_2 
\end{pmatrix} 
\simeq 
\begin{pmatrix}
 1 \\ 
 \sqrt{2} \eps^2 
\end{pmatrix}, 
\quad 
\Ykr{1}{4_2} = 
\begin{pmatrix}
 Y_3 \\ Y_4 \\ Y_5 \\ Y_6 \\
\end{pmatrix}
\sim \frac{1}{3} 
\begin{pmatrix}
\sqrt{2}\eps^4 \\ -3 \\ -3\sqrt{2} \eps \\ 2\eps^3 
\end{pmatrix}, 
\end{align}
at $k=1$. 
For reference, we show the expansions of the modular forms 
of up to the three dimensional representations for $k\le 6$. 
At $k=2$, 
\begin{align}
 Y^{(2)}_{1^1_2} \sim -\sqrt{2}\eps, 
\quad 
 Y^{(2)}_{2_0} \sim \frac{1}{3\sqrt{2}}
\begin{pmatrix}
 -3 \\ 8 \eps^3 
\end{pmatrix}, 
\quad 
\Ykr{2}{3_0} = 
\begin{pmatrix}
-1 \\ 2\eps^2 \\ 2\eps^4 
\end{pmatrix}. 
\end{align}
At $k=3$, 
\begin{align}
 \Ykr{3}{2^0_0} \sim &\ \frac{1}{\sqrt{3}}
\begin{pmatrix}
-1 \\ 3\sqrt{2}\eps^2 
\end{pmatrix}, 
\quad 
 \Ykr{3}{2^0_2} \sim 
\frac{1}{\sqrt{3}}
\begin{pmatrix}
6\eps^4 \\ -\sqrt{2} 
\end{pmatrix}, 
\quad 
 \Ykr{3}{2^1_2} \sim - \sqrt{2} 
\begin{pmatrix}
\eps \\ \sqrt{2}\eps^3 
\end{pmatrix}. 
\end{align}
At $k=4$, 
\begin{align}
\Ykr{4}{1^0_0} \sim &\ \frac{1}{\sqrt{3}}, \quad 
\Ykr{4}{1^0_1} \sim 2\eps^2, \quad 
\Ykr{4}{2_0} \sim  -\frac{1}{6\sqrt{2}}
\begin{pmatrix}
 3 \\ 16\eps^3 
\end{pmatrix}, \quad  
\Ykr{4}{2_2} \sim  -\frac{1}{3}
\begin{pmatrix}
 8\eps^4 \\ 3\eps 
\end{pmatrix},  \\ \notag 
\Ykr{4}{3^0} \sim &\ \sqrt{\frac{2}{3}}
\begin{pmatrix}
1 \\ 2\eps^2 \\ 6 \eps^4 
\end{pmatrix}, 
\quad 
\Ykr{4}{3^1} \sim  \sqrt{2}
\begin{pmatrix}
 -2\eps^3 \\ -2\eps^5 \\ \eps 
\end{pmatrix}.
\end{align}
At $k=5$,
\begin{align}
\Ykr{5}{2^0_0} \sim&\   \frac{1}{\sqrt{3}}
\begin{pmatrix}
 1 \\ \sqrt{2}\eps^2 
\end{pmatrix}, 
\quad 
\Ykr{5}{2^0_1} \sim
\begin{pmatrix}
 2\eps^2 \\ 2\sqrt{2}\eps^4 
\end{pmatrix}, 
\quad 
\Ykr{5}{2^0_2} \sim -\frac{1}{3}
\begin{pmatrix}
10\eps^4 \\ \sqrt{2} 
\end{pmatrix}, 
\\ \notag 
\Ykr{5}{2^1_1} \sim&\ \frac{1}{\sqrt{3}}
\begin{pmatrix}
-6\sqrt{2}\eps^5 \\ 2\eps 
\end{pmatrix}, 
\quad 
\Ykr{5}{2^1_2} \sim \frac{1}{\sqrt{3}}
\begin{pmatrix}
\sqrt{2}\eps \\ -6\eps^3 
\end{pmatrix}.   
\end{align}
At $k=6$, 
\begin{align}
 \Ykr{6}{1^0_0}\sim&\ \frac{1}{4\sqrt{2}}, 
\quad
 \Ykr{6}{1^1_0}\sim -2\sqrt{2}\eps^3,
\quad 
 \Ykr{6}{1^1_2} \sim -\sqrt{\frac{2}{3}}\eps, 
 \\ \notag 
 \Ykr{6}{2_0} \sim&\ \frac{1}{3\sqrt{6}}
\begin{pmatrix}
 -3 \\ 8\eps^3 
\end{pmatrix}, 
\quad 
 \Ykr{6}{2_1} \sim \frac{1}{3} 
\begin{pmatrix}
3\sqrt{2}\eps^2 \\ -8\sqrt{2}\eps^5 
\end{pmatrix}, 
\quad 
\Ykr{6}{2_2}\sim \frac{1}{6} 
\begin{pmatrix}
 16\eps^4 \\ -3\eps 
\end{pmatrix}, 
\\ \notag 
\Ykr{6}{3^0,1} \sim&\ \frac{1}{\sqrt{3}}
\begin{pmatrix}
 -1 \\ 2\eps^2 \\ 2\eps^4 
\end{pmatrix},
\quad 
\Ykr{6}{3^0,2} \sim \frac{1}{\sqrt{3}} 
\begin{pmatrix}
 0 \\ 4\eps^2 \\ -8\eps^4
\end{pmatrix}, 
\quad 
\Ykr{6}{3^1} \sim \frac{1}{\sqrt{3}}
\begin{pmatrix}
 -4\eps^3 \\ -12\eps^5 \\ -2\eps 
\end{pmatrix}.
\end{align}

{\small
\bibliography{ref_modular} 
\bibliographystyle{JHEP} 
}

\end{document}